\documentclass[a4paper]{jpconf}
\usepackage{graphicx,iopams,amsmath,cite}
\begin{document}
\title{Comparison of Hugoniots calculated for aluminum in~the framework of three quantum-statistical models}

\author{M~A~Kadatskiy and K~V~Khishchenko}

\address{Joint Institute for High Temperatures of the Russian Acad\-emy of Sciences, Izhorskaya 13 Bldg 2, Moscow 125412, Russia}

\ead{makkad@yandex.ru}

\begin{abstract}
The results of calculations of thermodynamic properties of aluminum under shock compression in the framework of the Thomas--Fermi model, the Thomas--Fermi model with quantum and exchange corrections and the Hartree--Fock--Slater model are presented. The influences of the thermal motion and the interaction of ions are taken into account in the framework of three models: the ideal gas, the one-component plasma and the charged hard spheres. Calculations are performed in the pressure range from 1 to $10^7$ GPa. Calculated Hugoniots are compared with available experimental data.
\end{abstract}



\section{Introduction}
Various quantum-statistical cell models with the approximation of self-consistent field are widely used for development of equations of state \cite{Bushman:1983, Lomonosov-LPB-2007, Khishchenko-2015}. The application of these approximations is even more valid, the higher the temperature and the density of matter \cite{Bushman:1983}.

The simplest of these models is the generalized Thomas--Fermi (TF) model \cite{Feynman:1949}, which is based on the semiclassical approximation for electrons and the Fermi--Dirac statistics. Considering of the exchange and the quantum effects within the framework of the TF approximation leads to the Thomas--Fermi with corrections (TFC) model \cite{Kirzhnits:1957}. Taking into account shell effects in the framework of semiclassical approximation is also workable \cite{Shpatakovskaya:1985, Shpatakovskaya:2012}. The separation of electrons into continuous and discrete energy spectrum and taking into account the exchange energy in the semiclassical approximation gives the equations of the Hartree--Fock--Slater (HFS) model \cite{Nikiforov_Novikov_Uvarov:2005}.

In addition to thermodynamic functions of electrons, which are derived from these models, it is also necessary to take into account for the component responsible for the thermal motion of the ions. At sufficiently high temperature, where electrons give the main contribution to the thermodynamics of substances, the condition of additivity is assumed valid, and it is possible to consider the contribution of ions as an additional term to the thermodynamic functions of electrons. 

The basic model of the accounting the thermal motion of ions is the ideal gas (IG) model \cite{Landau-Lifshitz-V-1980-eng}. By adding the correction for ion--ion interaction to the energy of the IG, it is possible to obtain the one-component plasma (OCP) model \cite{Hansen:1973}. The charged hard spheres (CHS) model takes into account an additional term to the ion pressure in connection with the finite size of ions \cite{Nikiforov_Novikov_Uvarov:2005}.
 
The physical accuracy of each of these ion models is not incontrovertible, and these models make different contribution to the equation of state in the region of shock-wave experiments. Therefore, it is important to consider all options for taking into account the ion contribution, the obtained dependencies being compared with available experimental data. Less consistent approach is implemented in a case each quantum-statistical electron model is put into correspondence with that ionic component, which provides the most consistent with experimental data total result, or in a case models with different electronic and ionic contribution are compared.

In this paper, the principle shock adiabat (principle shock Hugoniot) of aluminum is calculated for all 9 combinations of the above considered 3 electron and 3 ion models. The calculations were performed in the pressure range from 1 to $10^7$ GPa, which corresponds approximately to the temperature range from $10^{-2}$ to $10^4$ eV. At high pressures and temperatures, the compression ratio $\sigma = \rho/\rho_{00}$ (where $\rho$ is the density of the substance, $\rho_{00}$ is the initial density of the sample) on the shock Hugoniot ceases to depend on the pressure and reaches the asymptotic value
\begin{equation}
 \sigma_\mathrm{lim} = \frac{\gamma+1}{\gamma-1}.
\notag
\end{equation}
Here $\gamma = C_{P}/C_{V}$ is the adiabatic index, $C_{P}$ is the specific heat at constant pressure, $C_{V}$ is the specific heat at constant volume. At high temperatures, both electron and ion gases are almost homogeneous and perfect, and $\gamma = 5/3$ \cite{Zeldovich_Raizer:1967}.

It should be noted that, at temperatures greater than $10^3$ eV, it is necessary to take into account the contribution of equilibrium radiation to the total thermodynamics of substances \cite{Fortov:2012}. However, in the paper, this effect was not taken into account. Relativistic effects are also not taken into account; these effects become important only at temperatures $T > 10^5$ eV \cite{Kirzhnits:1972}.

\section{Cell quantum-statistical models}
Only one spherical cell (with radius $r_0$) is considered in the framework of the cell models. The volume of this cell is assumed equal to a volume, which is attributable to an average of one atom in a substance (Wigner--Seitz cell).

In the considered approach, electrons and ions are taken into account by separate models. In general, it is impossible to implement an exact separation of electronic and ionic contribution to the equation of state for matter. But if one proceeds from the assumption that the most significant contribution to thermodynamics is provided by electrons, which are interacting with each other and with ions, then the ion--ion interaction and thermal motion of the ions can be considered as additional components to the obtained values of the electronic pressure $P_{e}=P_{e}(\rho,T)$ and the electronic energy $E_{e}=E_{e}(\rho,T)$.

For the total pressure $P$ and the specific energy $E$ (in SI units), following expressions are used:
\begin{equation}
 P = \frac{E_\mathrm{H}}{a^3_0}(P_e+P_i), \quad E = \frac{E_\mathrm{H}}{A m_u}(E_e+E_i),
\notag
\end{equation}
where $P_{i} = P_{i}(\rho, T)$ and $E_{i} = E_{i}(\rho, T)$ are the pressure and energy of ions (all the components $P_e$, $P_i$, $E_e$ and $E_i$ are expressed in atomic units); $E_\mathrm{H}$ is the Hartree energy, $a_{0}$ is the Bohr radius, $m_{u}$ is the atomic mass unit, $A$ is the atomic mass in $m_u$.

\subsection{Models of the electronic part}
In the framework of the approximation of self-consistent field, the atomic potential is averaged over the different positions of the nuclei. The average potential $V(\vec{r})$ satisfies Poisson equation
\begin{equation}
 \Delta{V(\vec{r})} = -4\pi{Z}\delta(\vec{r})+4\pi\rho_{e}(\vec{r}),
\notag
\end{equation}
where $Z$ is the nuclear charge, $\delta$ is the Dirac delta function, $\rho_{e}(\vec{r})$ is the electron density, $\mathrm{d}N(\vec{r}) = \rho_{e}(\vec{r})\mathrm{d}\vec{r}$ is the number of electrons in a small volume $\mathrm{d}\vec{r}$. Let us consider the plasma without the distinguished direction. Then the atomic potential is assumed spherically symmetric. By this way, the transition from vector to scalar functions is implemented.

Within the atomic cell, the electron states are divided into three groups: (i) states with a continuous energy spectrum (with a corresponding electron density $\rho_\mathrm{cont}$), (ii) states with a discrete energy spectrum ($\rho_\mathrm{bound}$) and (iii) states of the intermediate group ($\rho_\mathrm{band}$). Accordingly, let us represent the electron density as a sum of three terms:
\begin{equation}
\rho_{e}(r) = \rho_\mathrm{cont}(r)+\rho_\mathrm{bound}(r)+\rho_\mathrm{band}(r).
\notag
\end{equation}
In addition, there is the charge neutrality condition for the atomic cell:
\begin{equation}
4\pi\int^{r_0}_{0}\rho_{e}(r)r^2dr = Z.
\notag
\end{equation}

In the framework of the average atom approximation, instead of a set of ions in different states, one atom with average occupation numbers is considered. Occupation numbers themselves are calculated according to the Fermi--Dirac distribution.

It is possible to describe thermodynamics of matter at high density or high temperature successfully by using the TF model \cite{Feynman:1949}. In this model, all electrons belong to the continuous spectrum so $\rho_{e\/\mathrm{TF}}(r)=\rho_\mathrm{cont}(r)$. The basic equations of the TF model and the formulas for thermodynamic values are described in \cite{Feynman:1949}.

Taking into account exchange and quantum second-order corrections in $\hbar$, Kirzhnits implemented the transition from TF to TFC \cite{Kirzhnits:1957}. In \cite{Kalitkin:1960}, expressions for corrections to the thermodynamic functions were derived, in particular, for pressure and energy. The efficient numerical method for solving the equations of the TFC model was proposed in \cite{Kalitkin:1975}. To derive the thermodynamic functions of the TFC model, it is necessary to calculate the appropriate functions of the TF model:
\begin{equation}
P_{e\/\mathrm{TFC}} = P_{e\/\mathrm{TF}}+\Delta P_{e\/\mathrm{TFC}}, \quad E_{e\/\mathrm{TFC}} = E_{e\/\mathrm{TF}}+\Delta E_{e\/\mathrm{TFC}}.
\notag
\end{equation}
Values of corrections $\Delta P_{e\/\mathrm{TFC}}$ and $\Delta E_{e\/\mathrm{TFC}}$ should satisfy the condition of smallness in comparison with base values \cite{Kirzhnits:1957}.

Although the electron density of the TF and TFC models is consistent with the atomic potential $V(\vec{r})$, the wave functions calculated for this potential yield the electron density, which does not coincide with the original. This is because the TF and TFC models do not reflect oscillations of the wave functions.

In the HFS model \cite{Nikiforov_Novikov_Uvarov:2005}, these oscillations are effectively taken into account. It is provided by choosing an effective energy boundary of the continuous spectrum $\epsilon_0$, so that the electrons, which give the main contribution to the oscillating part of the electron density, belong to the discrete spectrum. The Coulomb contribution to the self-consistent potential of the HFS model is calculated similarly to other cell models. The density of the bound electrons (with $\rho_\mathrm{bound}$) is calculated directly in terms of the wave functions. These wave functions are solutions of the Schr\"{o}dinger equation with boundary conditions $R_{nl}(0) = 0$, $R_{nl}(r_0) = 0$ and the normalization condition $\int^{r_0}_{0}R_{nl}^2(r)\mathrm{d}r=1$ \cite{Nikiforov_Novikov_Uvarov:2005}. Due to the influence of neighboring atoms, the spectrum of the electrons of the intermediate group consists of bands of allowed energies. Periodic boundary conditions \cite{Perrot:1975} are used for taking into account the band structure. In the HFS model, in addition to the Coulomb contribution, there is an exchange correction, which is calculated in the semiclassical approximation \cite{Nikiforov_Novikov_Uvarov:2005}. Found value of the self-consistent potential of the HFS model is used to obtain thermodynamic functions of electrons, in particular, pressure and energy \cite{Nikiforov_Novikov_Uvarov:2005}.

\subsection{Models of the ionic part}
The IG model is the simplest approximation, which takes into account the thermal motion of the ions \cite{Landau-Lifshitz-V-1980-eng}. This approximation is applicable for heated rarefied matter. The pressure (in units of $E_\mathrm{H}/a_0^3$) and energy (in units of $E_\mathrm{H}$) of an ideal gas is expressed by
\begin{equation}
P_{i\/\mathrm{IG}} = \frac{3\theta}{4\pi r^3_0}, \quad E_{i\/\mathrm{IG}} = \frac{3\theta}{2}.
\notag
\end{equation}
In these formulae, $\theta$ is the electron temperature (in units of $E_\mathrm{H}$).

In contrast to the IG model, OCP model \cite{Hansen:1973} takes into account the interaction of ions. In the framework of this model, ions are considered as point particles with charge $Z_0 = (4/3)\pi r_0^3 \rho_e(r_0)$ moving in homogeneous neutralizing surroundings (of compensating charge $-Z_0$). In the OCP approximation,
\begin{equation}
P_{i\/\mathrm{OCP}} = \frac{3\theta}{4\pi r^3_0}\left(1+\frac{\Delta E_i}{3\theta}\right), \quad E_{i\/\mathrm{OCP}} = \frac{3\theta}{2}+\Delta E_i,
\notag
\end{equation}
where $\Delta E_i$ is the correction to the internal energy in the OCP model \cite{Hansen:1973}:
\begin{gather}
 \Delta E_i = \left\{
 \begin{array}{cl}
 \Delta E_\mathrm{OCP}, & \mbox{if~} \Delta E_\mathrm{OCP}<3\theta/2, \\
 3\theta/2, & \mbox{if~} \Delta E_\mathrm{OCP}\geqslant 3\theta/2,
 \end{array} \right.
		\nonumber
		\\
		\Delta E_\mathrm{OCP}=\theta \Gamma^{3/2}_e \sum^4_{i=1}\frac{a_i}{(b_i+\Gamma_e)^{i/2}} -\theta\Gamma_e a_1, \quad \Gamma_e = \frac{Z_0^2}{\theta r_0},
		\nonumber
\\
a_1=-0.895929, \quad
a_2=0.11340656, \quad
a_3=-0.90872827, \quad
a_4=-0.11614773, 
\notag
\\
b_1= 4.666486, \quad
b_2= 13.675411, \quad
b_3= 1.8905603, \quad
b_4= 1.0277554. 
\notag
\end{gather}

The CHS model takes into account the strong repulsion between ions at distance $r_{\ast}$, which is equal to the effective radius of the ion core \cite{Nikiforov_Novikov_Uvarov:2005}. The value of $r_{\ast}$ is calculated by formula
\begin{equation}
 4\pi\int^{r_{\ast}}_{0}\rho_e(r) r^2 \mathrm{d}r = Z-Z_0.
\notag
\end{equation}
The expression for the pressure is obtained by adding the correction of the CHS model to pressure of OCP model, and energy of CHS model is the same as energy of OCP model, in particular:
\begin{equation}
P_{i\/\mathrm{CHS}} = \frac{3\theta}{4\pi r^3_0}\left(1+\frac{2\eta(2+\eta)}{(1-\eta)^2}+\frac{\Delta E_i}{3\theta}\right), \quad E_{i\/\mathrm{CHS}} = E_{i\/\mathrm{OCP}}=\frac{3\theta}{2}+\Delta E_i,
\notag
\end{equation}
where $\eta = (r_{\ast}/r_0)^3$ is the packing parameter.

\section{Calculation of Hugoniots}
The choice of aluminum for comparison of models is due to the availability of numerous experimental data in the wide range of temperatures and pressures for this metal. The nucleus charge of aluminum is relatively small ($Z = 13$), so it is possible to use non-relativistic equations without noticeable loss of accuracy \cite{Nikiforov_Novikov_Uvarov:2005}.

Under compression of a substance across the shock-wave front, the Hugoniot relation holds \cite{Zeldovich_Raizer:1967}:
\begin{equation}
E = E_0+\frac{1}{2}(P+P_0)\left(\frac{1}{\rho_{00}}-\frac{1}{\rho}\right),
\notag
\end{equation}
where $P$, $E$ and $\rho$ are the pressure, specific internal energy and density of the shock-compressed matter; $P_0$, $E_0$ and $\rho_{00}$ are relevant parameters in the initial state before the front.

In this paper, the values of $P_0$, $E_0$ and $\rho_{00}$ are selected from the experimental data. The case is considered of solid samples with $\rho_{00}$ = $\rho_{0}$, where $\rho_{0}$ is the density such that $E = E_0$ at $P = P_0$ and $\rho = \rho_0$. For aluminum, $E_0 = -12.1$ kJ/g and $\rho_0 = 2.712$ g/cm$^3$ at $P_0 = 0.1$ MPa.

\begin{figure}
\begin{center}
\includegraphics[width=0.9\columnwidth]{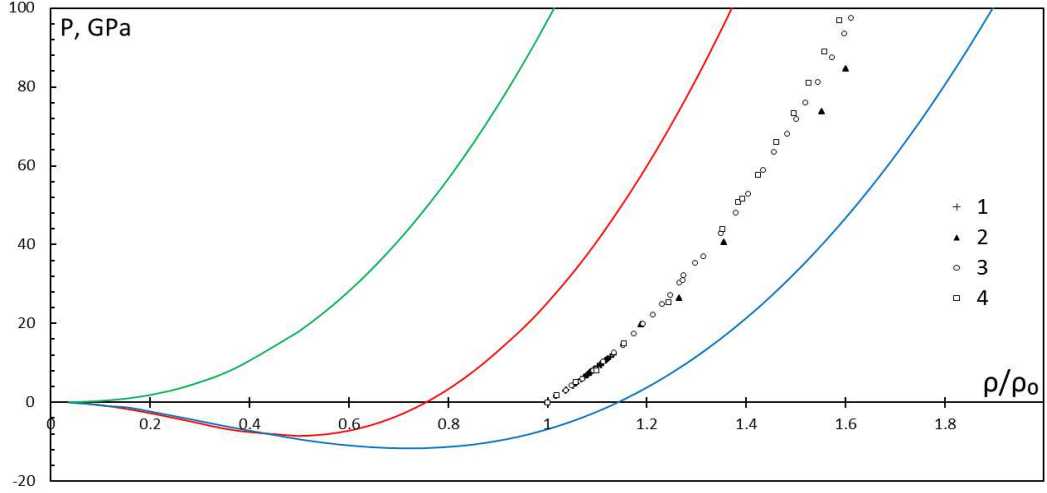}
\hfill
\end{center}
\caption{The pressure in aluminum at the room temperature $T = 293$ K according to different models of electronic part as a function of the compression ratio $\sigma=\rho/\rho_0$. The ionic contribution is taken into account by CHS model. Green curve---TF. Blue curve---TFC. Red curve---HFS. Experimental data: 1---\!\!\cite{Syassen:1978}, 2---\!\!\cite{Greene:1994}, 3---\!\!\cite{Dewaele:2004}, 4---\!\!\cite{Akahama:2006}.}
\label{fig1}
\end{figure}

The cell approximation is too inaccurate in the domain close to normal conditions for temperature ($T=T_0=293$ K) and pressure ($P=P_0 = 0.1$ MPa). In particular, for models I = TF, TFC and HFS the computed theoretical value of the normal density $\rho_{0\mathrm{I}}$, which is determined by the condition $P_0 = P(\rho_{0\mathrm{I}}, T_0)$, differs from the experimental value $\rho_{0}$ \cite{Syassen:1978, Greene:1994, Dewaele:2004, Akahama:2006}. Isotherms $T = T_0$ of aluminum, calculated by different models, are shown in figure~\ref{fig1}. At this temperature, the contribution of the ions is negligible, so the results do not depend on the model of the ionic part. Comparison of the isotherms shows that the results of the HFS computations occupy an intermediate position between the results of the TF and TFC models. Moreover, the condensed states of matter are not implemented in TF, and the estimated value $\rho_{0\mathrm{TF}}$ close to zero. The TFC model implements overestimated value of the normal density of aluminum $\rho_{0\mathrm{TFC}} = 1.416\rho_{0} = 3.096$~g/cm$^3$, and the HFS model underestimates value of the normal density $\rho_{0\mathrm{HFS}} = 0.7552\rho_{0} = 2.048$~g/cm$^3$. These differences have an effect on the behavior of the shock Hugoniots at its initial section ($P < 0.5$~TPa).

\begin{figure}[t]
\begin{center}
\includegraphics[width=0.9\columnwidth]{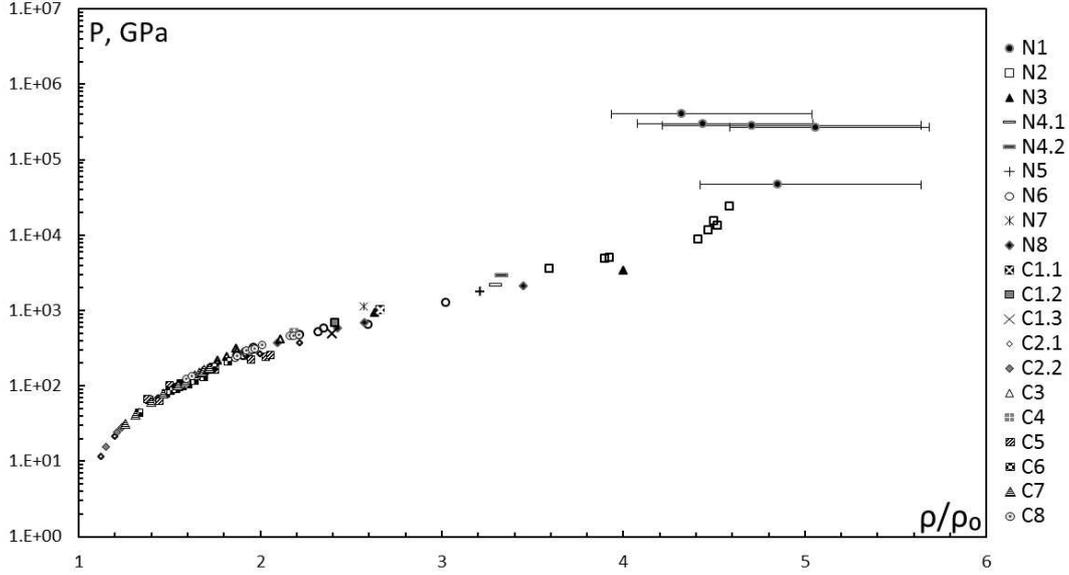}
\hfill
\end{center}
\caption{Experimental data on the shock compressibility of aluminum. The experiments with attainable pressure less than 1 TPa: C1.1---\!\!\cite{Trunin_c1.1:1995}, C1.2---\!\!\cite{Trunin_c1.2:1995}, C1.3---\!\!\cite{Trunin_c1.3:1986}, C2.1---\!\!\cite{Al'tshuler_c2.1:1974}, C2.2---\!\!\cite{Al'tshuler_c2.2:1974}, C3---\!\!\cite{Glushak_c3:1989}, C4---\!\!\cite{Kormer_c4:1962}, C5---\!\!\cite{Skidmore_c5:1962}, C6---\!\!\cite{Isbell_c6:1968}, C7---\!\!\cite{Mitchell_c7:1981}, C8---\!\!\cite{Knudson_c8:2003}. The experiments with attainable pressure higher than 1 TPa: N1---\!\!\cite{Vladimirov_N1:1984}, N2---\!\!\cite{Avrorin_N2:1986}, N3---\!\!\cite{Simonenko_N3:1985}, N4.1---\!\!\cite{Ragan_N4.1:1982}, N4.2---\!\!\cite{Ragan_N4.2:1984}, N5---\!\!\cite{Podurets_N5:1994}, N6---\!\!\cite{Trunin_N6:1995}, N7---\!\!\cite{Volkov_N7:1980}, N8---\!\!\cite{Al'tshuler_N8:1977}.}
\label{fig2}
\end{figure}

The experimental data of shock compression of solid aluminum are shown in figure~\ref{fig2}. The main set of experimental points belongs to the range of pressures $P < 1$ TPa \cite{Trunin_c1.1:1995, Trunin_c1.2:1995, Trunin_c1.3:1986, Al'tshuler_c2.1:1974, Al'tshuler_c2.2:1974, Glushak_c3:1989, Kormer_c4:1962, Skidmore_c5:1962, Isbell_c6:1968, Mitchell_c7:1981, Knudson_c8:2003}. In this range, the experimental Hugoniot of aluminum is determined quite reliably. Pressure range $P > 1$ TPa corresponds to data mainly produced by experiments with nuclear explosions \cite{Vladimirov_N1:1984, Avrorin_N2:1986, Simonenko_N3:1985, Ragan_N4.1:1982, Ragan_N4.2:1984, Podurets_N5:1994, Trunin_N6:1995, Volkov_N7:1980, Al'tshuler_N8:1977}. These experiments have a relatively large error in determination of the density of samples.

\begin{figure}[p]
	\begin{center}
		\begin{minipage}{0.63\linewidth} 
			\includegraphics[width=1\columnwidth]{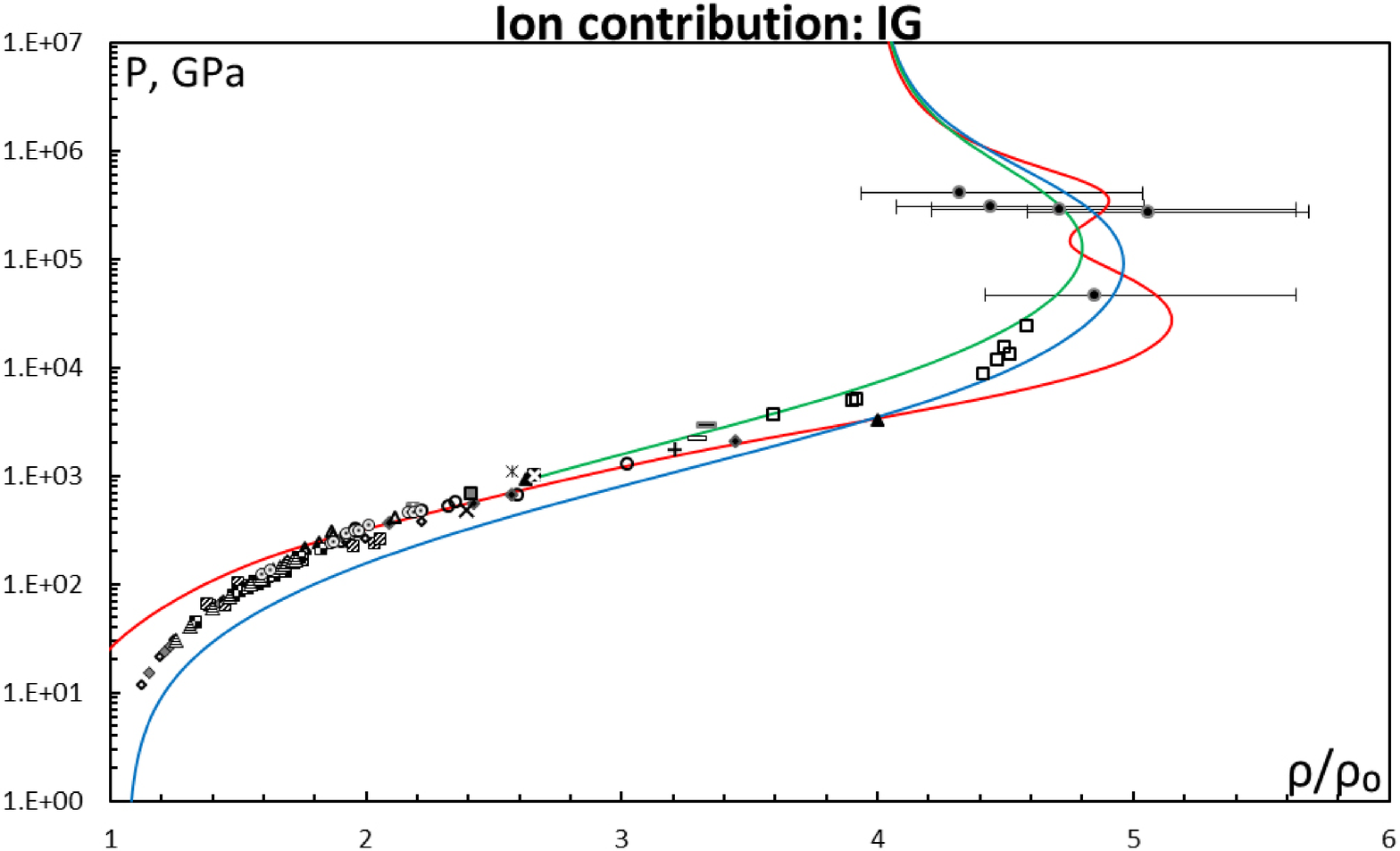}
			\includegraphics[width=1\columnwidth]{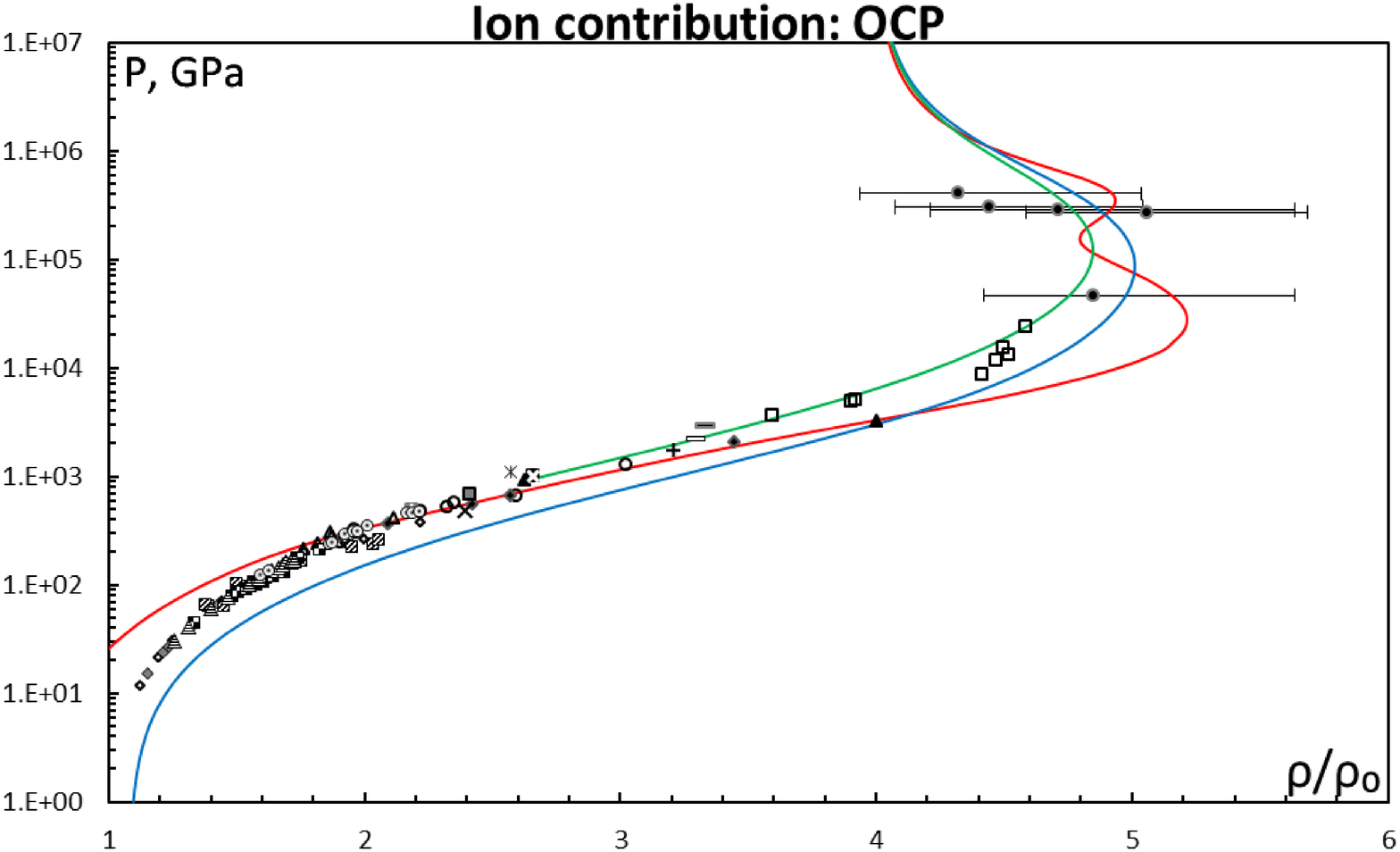}
			\includegraphics[width=1\columnwidth]{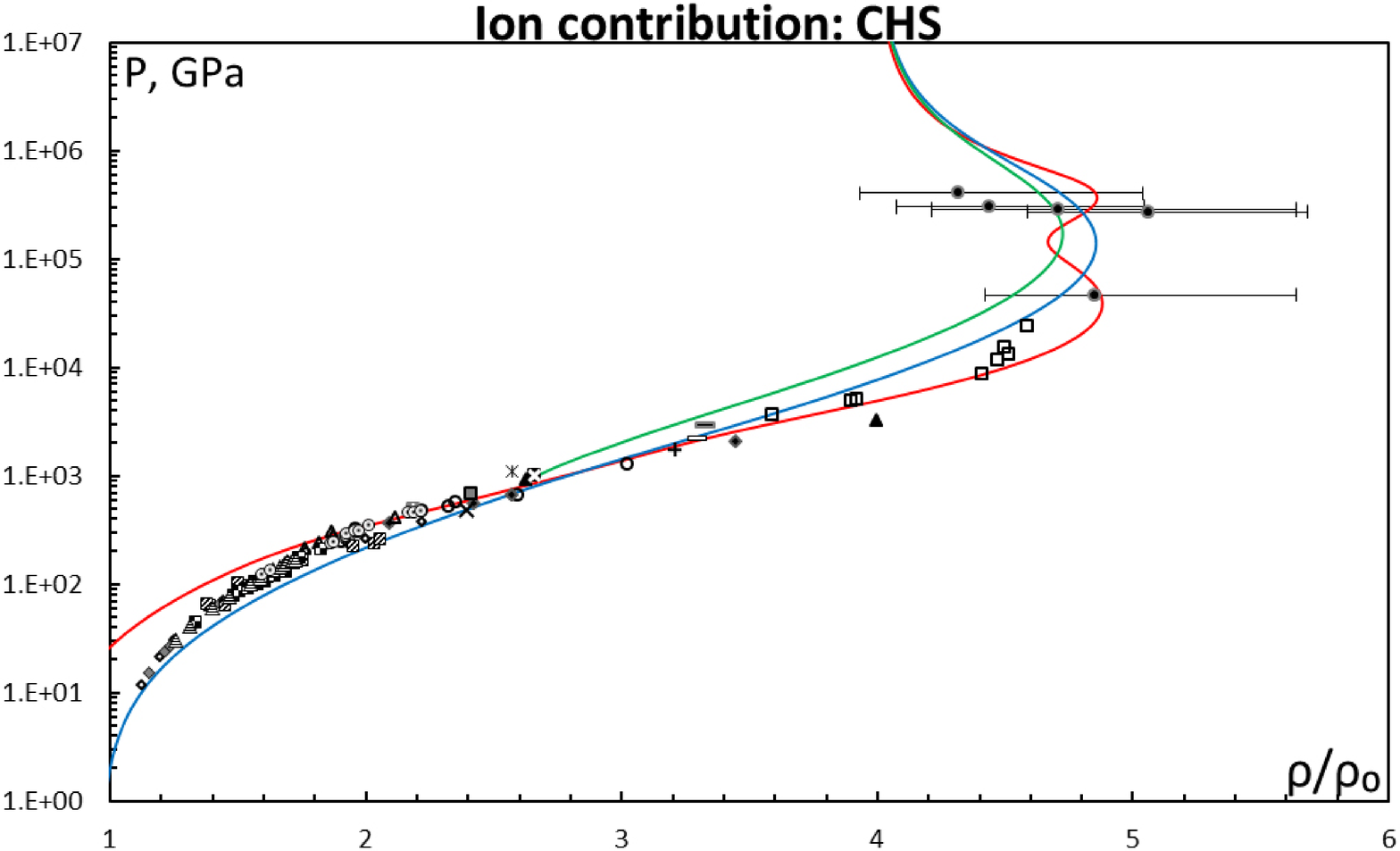}
		\end{minipage}
	\end{center}
\caption{Shock Hugoniot of aluminum for different ways of taking into account the electronic contribution to the equation of state (green curve---TF, blue curve---TFC, red curve---HFS). Separate graphs show the results for various ionic models. Experimental data: see figure~\ref{fig2}.}
\label{fig3}
\end{figure}

\begin{figure}[p]
	\begin{center}
		\begin{minipage}{0.63\linewidth} 
			\includegraphics[width=1\columnwidth]{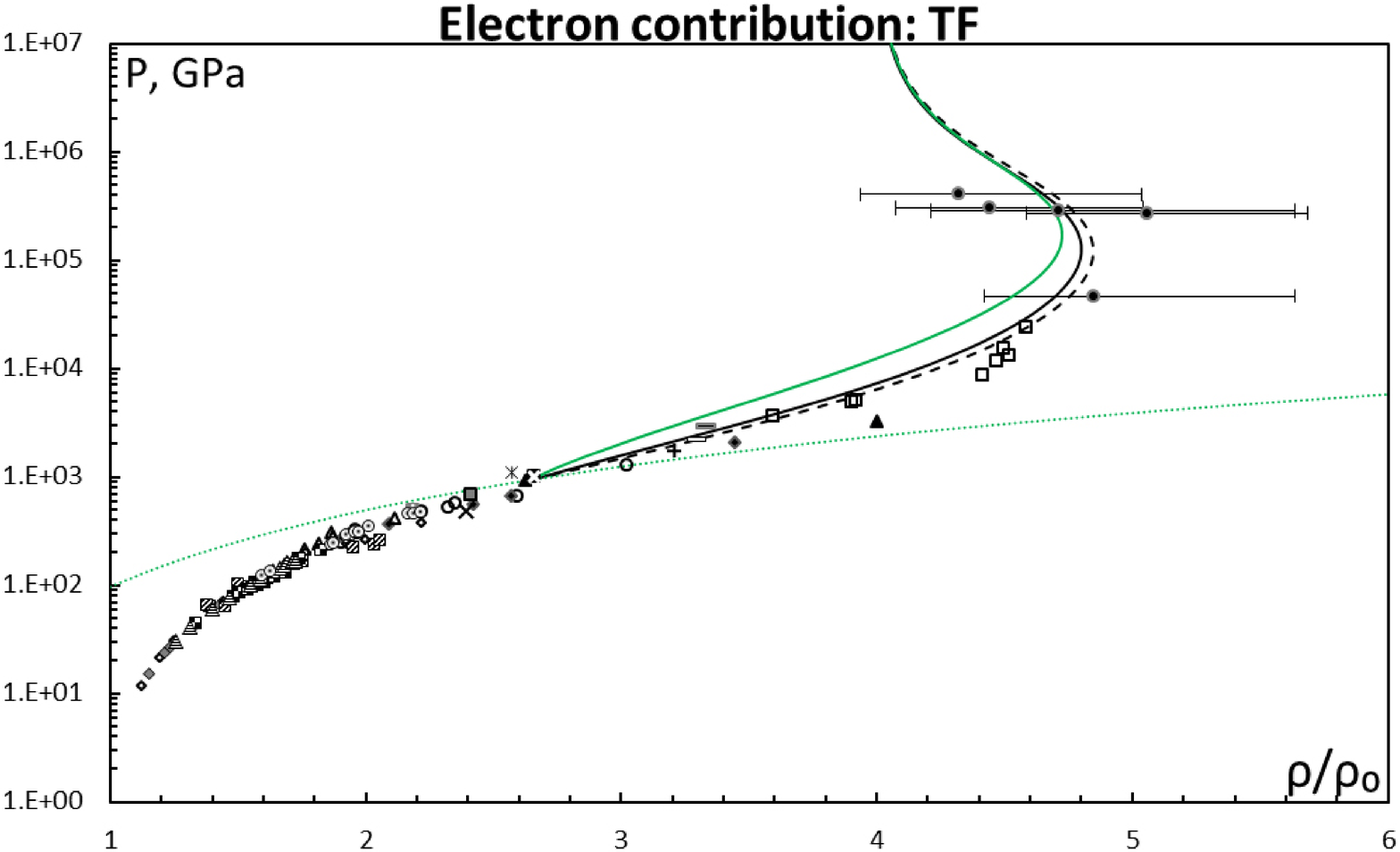}
			\includegraphics[width=1\columnwidth]{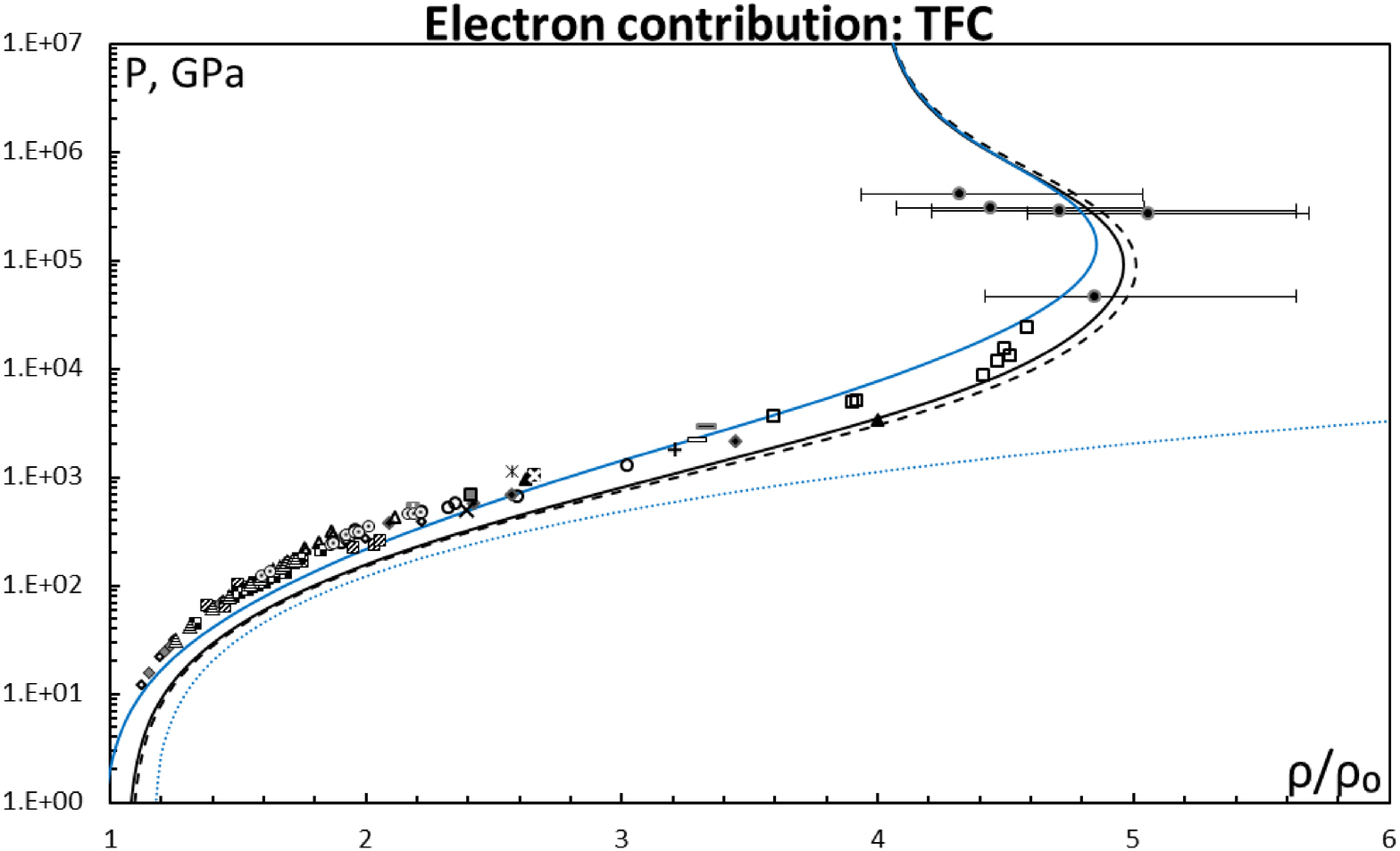}
			\includegraphics[width=1\columnwidth]{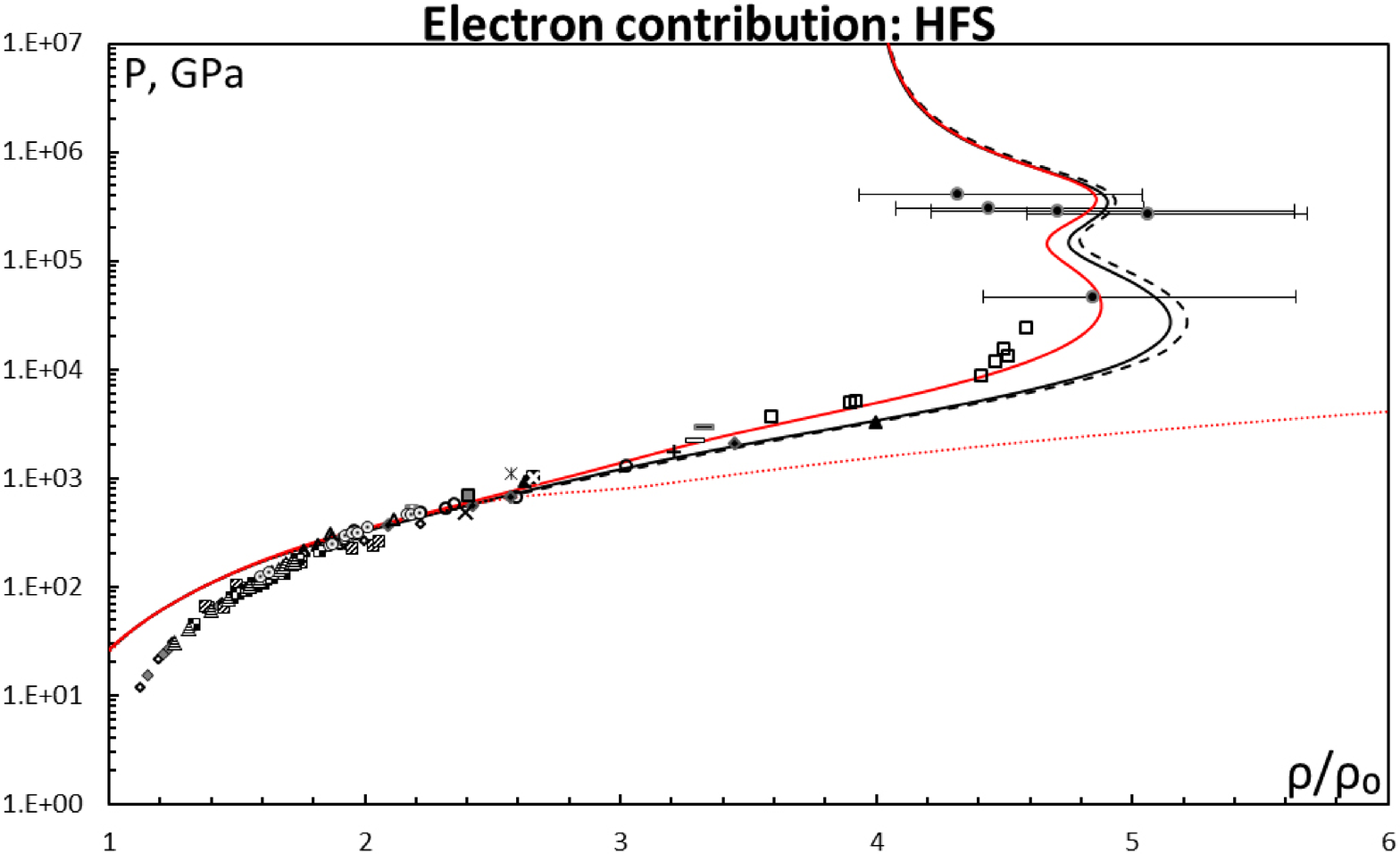}
		\end{minipage}
	\end{center}
\caption{Shock Hugoniot of aluminum for different ways of taking into account the ionic contribution to the equation of state (black solid curve---IG, black dashed curve---OCP, colored solid curve---CHS). Also shown is isotherm $T = 0$ K (colored dotted curve). Separate graphs show the results for various electronic models. Experimental data: see figure~\ref{fig2}.}
\label{fig4}
\end{figure}

Hugoniots for various combinations of electronic and ionic parts are shown in figure~\ref{fig3} and~\ref{fig4}. Figure~\ref{fig4} also shows the isotherm $T = 0$ K.

\begin{figure}[p]
	\begin{center}
		\begin{minipage}{0.63\linewidth}
			\includegraphics[width=1\columnwidth, keepaspectratio]{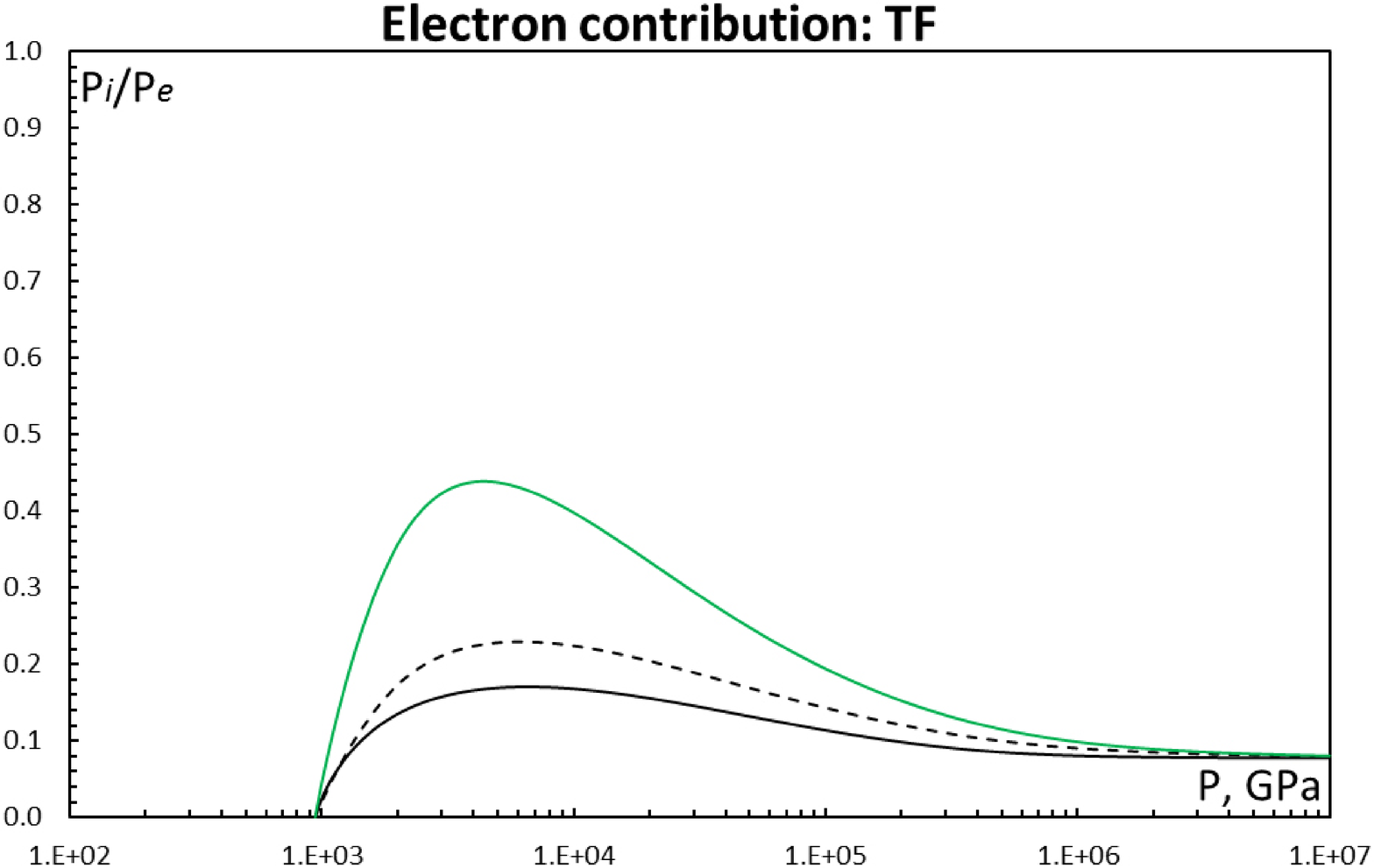}
			\includegraphics[width=1\columnwidth, keepaspectratio]{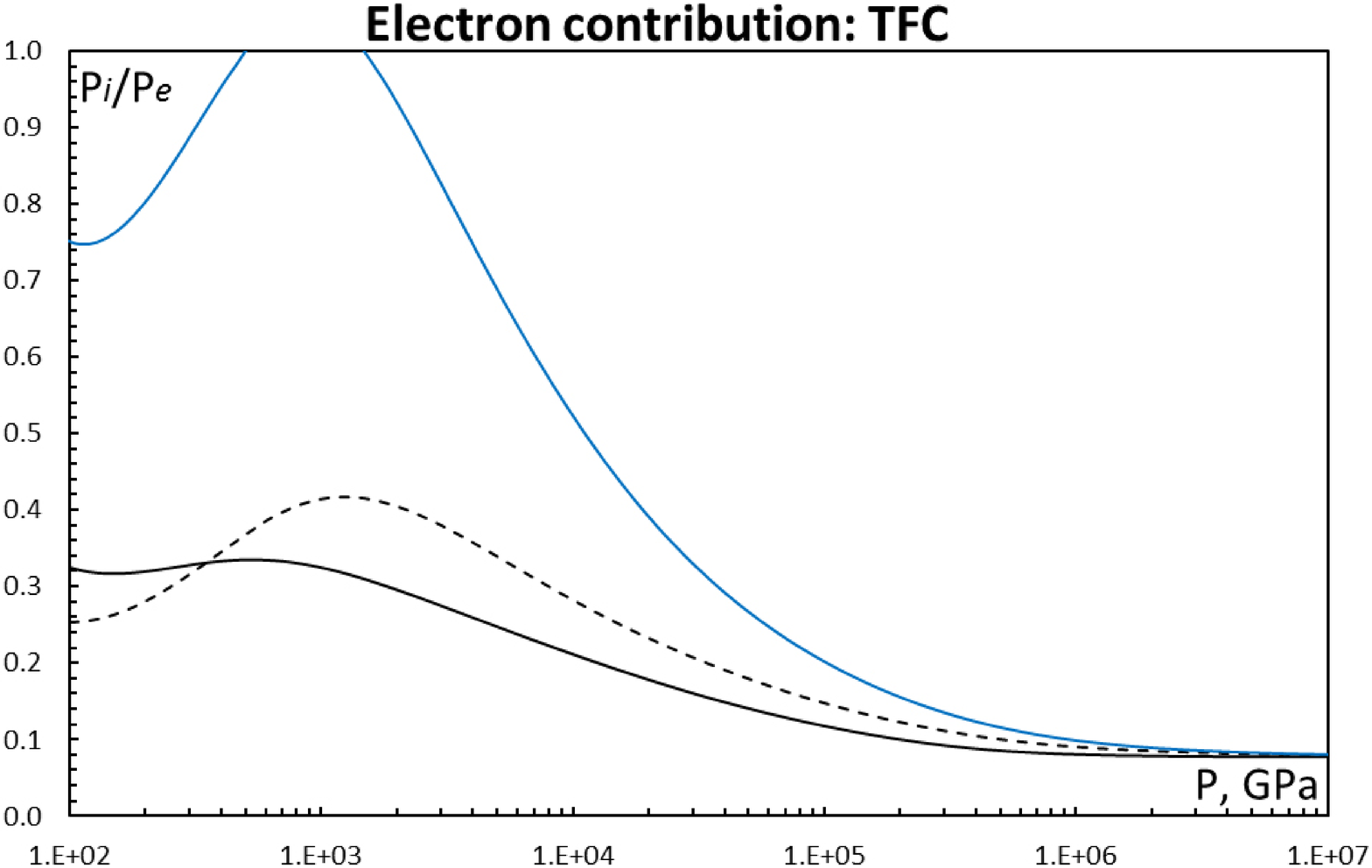}
			\includegraphics[width=1\columnwidth, keepaspectratio]{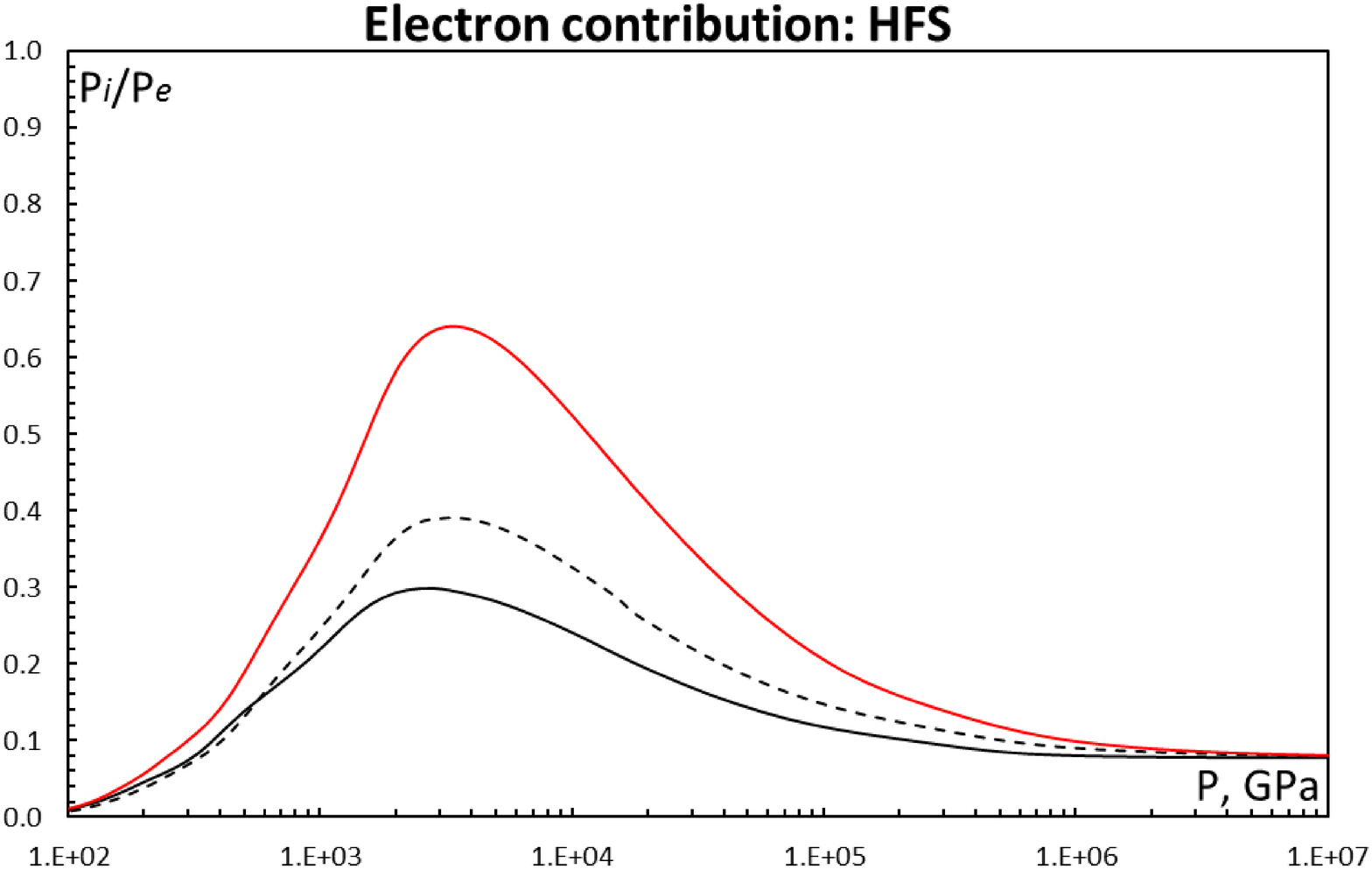}
		\end{minipage}
	\end{center}
	\caption{Ratio of the ion pressure to the electron pressure on the shock Hugoniot of aluminum, according to different models of the ionic contribution (black solid curve---IG, dashed curve---OCP, colored solid curve---CHS). Separate graphs shows the results for various electronic models.}
\label{fig5}
\end{figure}

At pressures $P \approx 1\textup{--}500$ TPa, with increasing pressure and temperature, the choice of model of the ionic part has a significant impact on the behavior of shock Hugoniot (see figure~\ref{fig4}). In this range, the main inaccuracy is connected with lack of physical reliability of the IG, OCP and CHS models. In addition, in this range thermodynamic values of ions reach magnitudes comparable to electronic terms (see figure~\ref{fig5}). Violation of the additivity conditions may lead to noticeable inaccuracy.

At pressures $P > 500$ TPa, the compression ratio $\sigma$ is changed from the maximum achieved value to the limit $\sigma_\mathrm{lim} = 4$. In this range, the ionic contribution is decreased compared with the electronic contribution, and distinctions between different ionic models are leveled (the nonideality parameter $\Gamma_e \rightarrow 0$). In this region, main inaccuracy of the models is connected with the unaccounted influence of the equilibrium thermal radiation on the thermodynamics of substances \cite{Fortov:2012}.

The qualitative difference in the behavior of the TFC and HFS models in the region of strong loading is due to the fact that in the framework of the TFC model shell effects are not taken into account \cite{Shpatakovskaya:1985, Shpatakovskaya:2012}. Oscillatory behavior of the shock Hugoniots calculated by the HFS model is connected with the ionization of electron shells with increasing temperature.

In the considered shock-wave experiments, the density of matter was not measured directly. By using the Hugoniot equation, the density can be expressed in terms of mass velocity $U$ and the wave velocity $D$ \cite{Zeldovich_Raizer:1967} as
\begin{equation}
\rho=\frac{D\rho_{00}}{D-U}, \quad U=\frac{P-P_0}{D\rho_{00}}.
\notag
\end{equation}

\begin{figure}[p]
	\begin{center}
		\begin{minipage}{0.63\linewidth}
			\includegraphics[width=1\columnwidth, keepaspectratio]{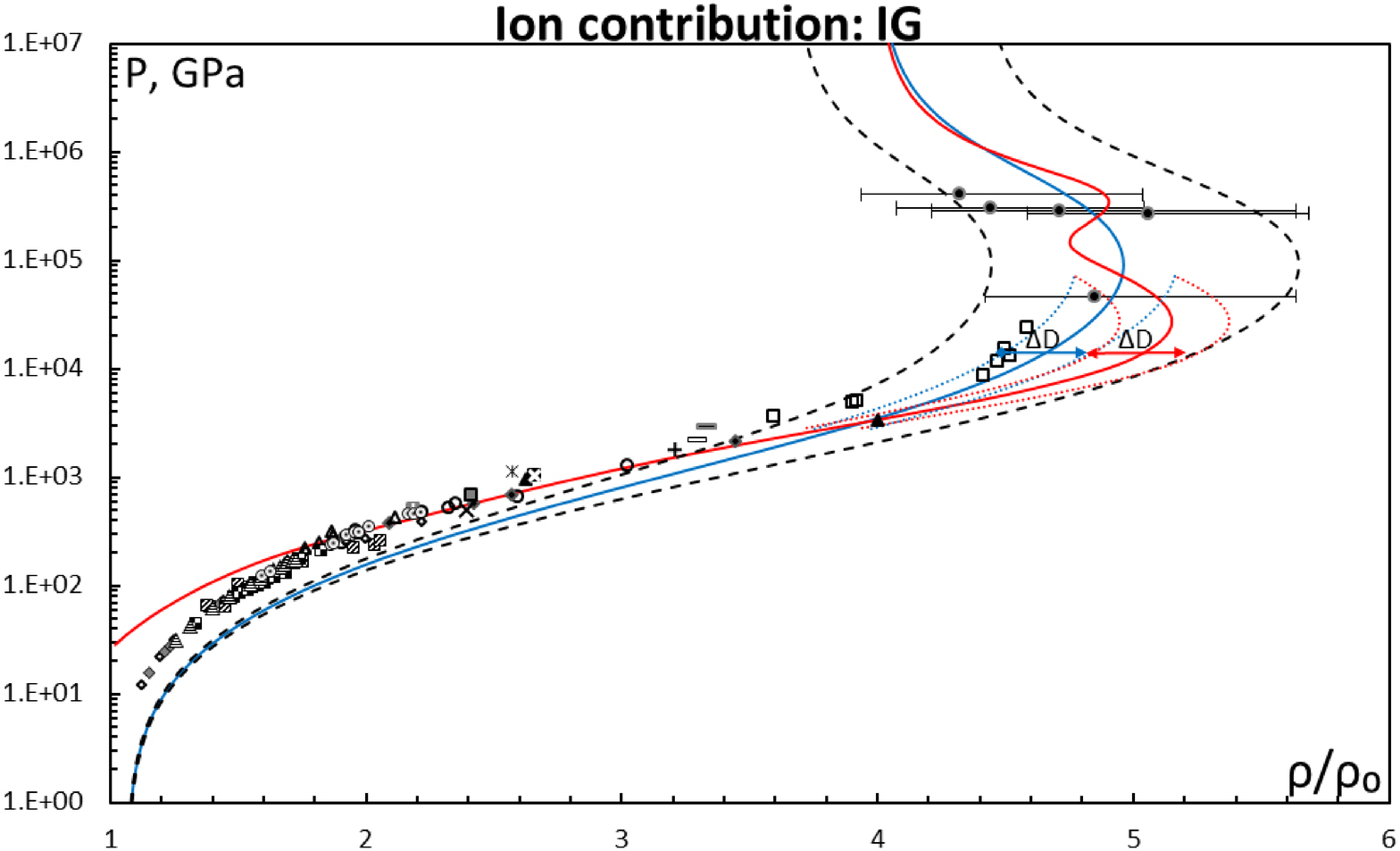}
			\includegraphics[width=1\columnwidth, keepaspectratio]{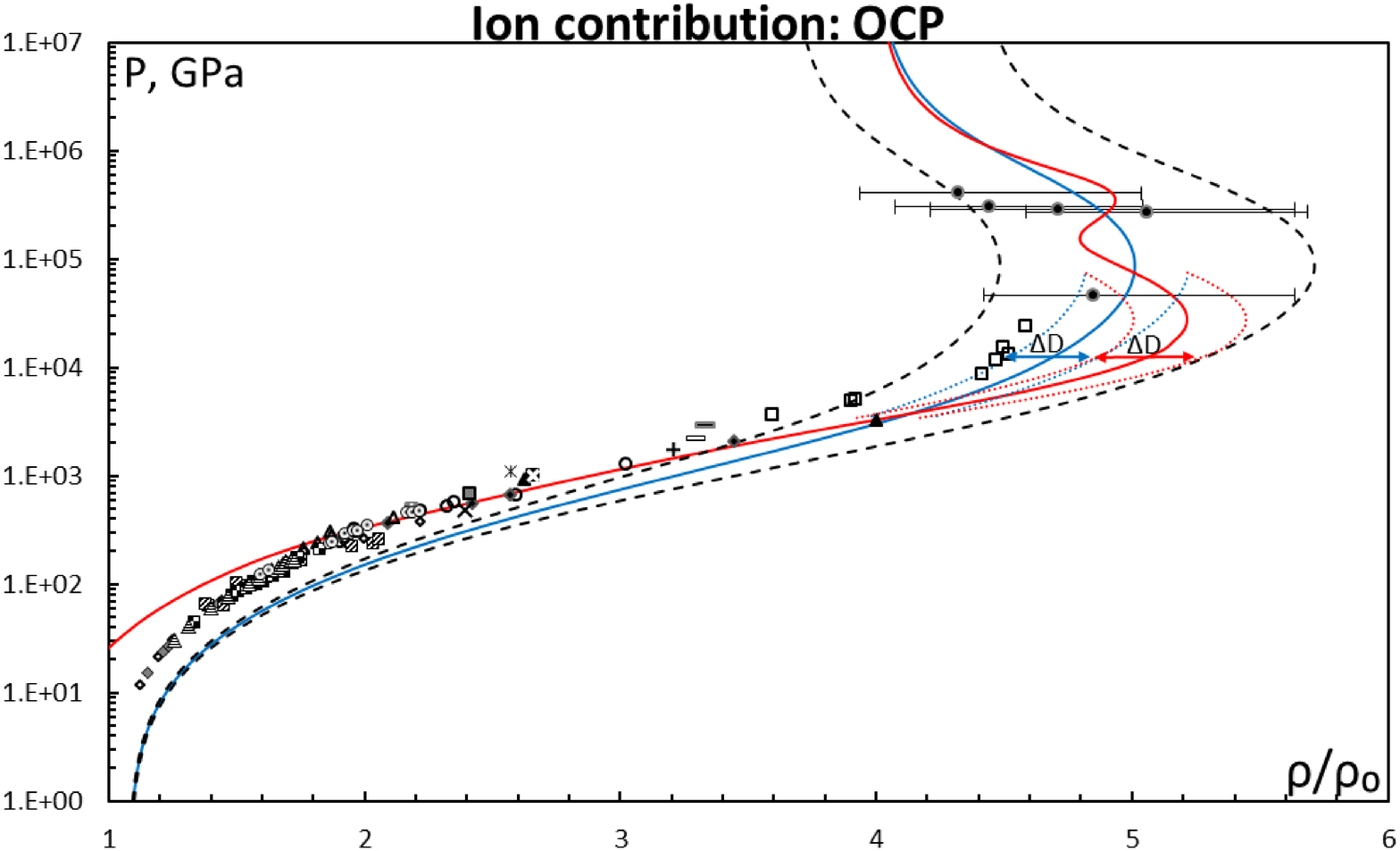}
			\includegraphics[width=1\columnwidth, keepaspectratio]{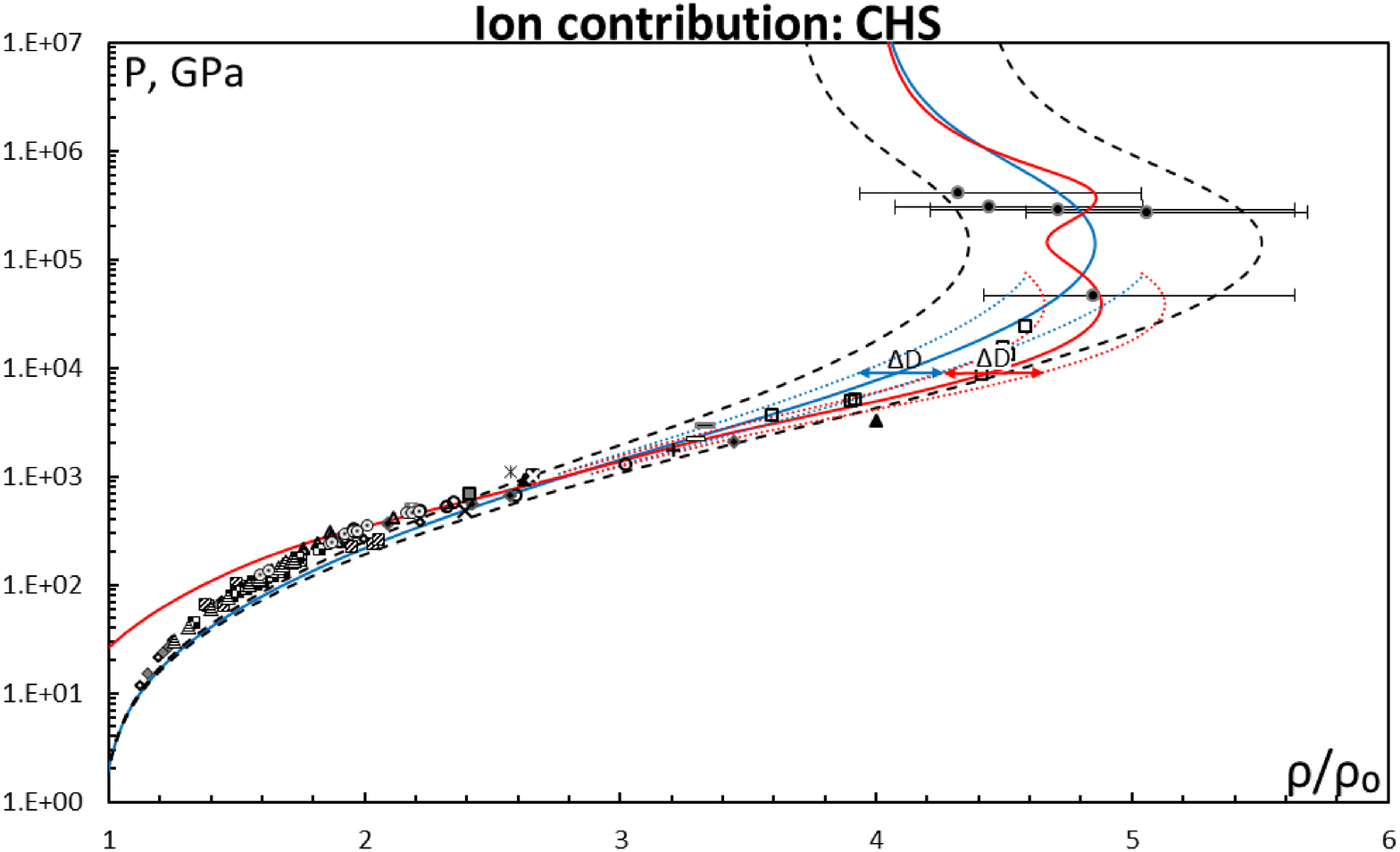}
		\end{minipage}
				\end{center}
	\caption{Shock Hugoniots of aluminum calculated by the TFC (blue solid curve) and HFS (red solid curve) models. The black dashed curve is connected with deviation of the position of the Hugoniots calculated by the TFC model in the case the value of $D$ is changed on $\pm1.5\%$. Colored dotted curves correspond to the positions where the value of the deviation of $D$ is $\Delta$: blue dotted curve---TFC, red dotted curve---HFS. Separate graphs show the results for various ionic models. In particular, for IG model---$\Delta = \pm0.5\%$, for OCP model---$\Delta = \pm0.5\%$, for CHS model---$\Delta = \pm0.6\%$. Experimental data: see figure~\ref{fig2}.} 
\label{fig6}
\end{figure}

There is the observed oscillation of the shock Hugoniot calculated by the HFS model relatively to curve calculated by the TFC model at pressures $P \approx 1\textup{--}500$ TPa. In this pressure range, even a small error in determining value of $D$ strongly affects the resulting value of $\rho$. In \cite{Avrorin_N2:1986}, the error in determining value of the wave velocity for a single experimental point is approximately $1.5\%$. In figure~\ref{fig6}, the boundaries of changing the position of the shock Hugoniot calculated by the TFC model where the deviation of $D$ is $1.5\%$ are marked. It is seen that the oscillation effects on the shock Hugoniot of aluminum do not exceed the limit at which these effects can be reliably detected experimentally with such accuracy. It is possible to make an estimate that, for the experimental detection of oscillation effects, the measurement of error of $D$ should be less than $0.5\%$.

\section{The influence of the choice of the effective boundary of the continuous spectrum}
In the HFS model, for selection the effective boundary of the continuous spectrum $\epsilon_0$, next condition is used:
\begin{equation}
\frac{8\sqrt{2}}{3\pi}\int^{r_0}_{0}\left(\max\left[0,\epsilon_0+V(r)\right]\right)^{3/2}r^2 \mathrm{d}r=\sum_{nl}2\left(2l+1\right)+\sum_{nlm}\int^{k_0}_{0}\frac{6k^{2} \mathrm{d} k}{k_0^3}.
\label{Eps0}
\end{equation}
Here, summation over the quantum numbers $n$, $l$, $m$ and integration over the quasi-momentum $k$ are carried out over states with energy $\epsilon_{nl}<\epsilon_0$, $\epsilon_{nlm}(k)<\epsilon_0$. This condition provides the thermodynamic consistency of the equations of the HFS model. But an ambiguity arises from the fact that the equation~(\ref{Eps0}) has multiple solutions.

In \cite{Nikiforov_Novikov_Uvarov:2005}, it was proposed to fix the number of the root $N$ of equation~(\ref{Eps0}) so that the value of $\epsilon_0$ at $T = 0$ K, $\rho = \rho_0$ belongs to the conduction band for metals. In the framework, for aluminum, this condition yields the value $N = 4$, close to the number of electron shells $nl$ in the atom. 

In this paper, for all calculations by the HFS model, $\epsilon_0$ was taken as the maximum value of the root with $N = N_\mathrm{max}$ of equation~(\ref{Eps0}) for a fixed main quantum number $n = 3$. This choice provides a smooth behavior of the thermodynamic dependencies throughout the computational domain and does not depend on other assumptions.

\begin{figure}[b]
	\begin{center}
			\includegraphics[width=0.9\columnwidth, keepaspectratio]{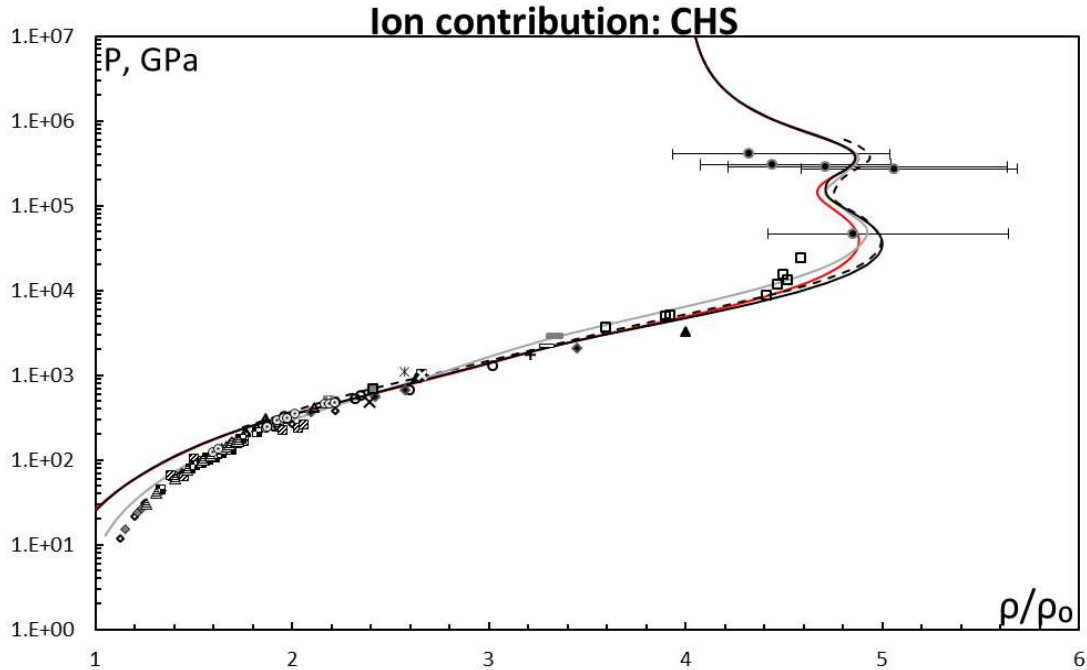}
				\end{center}
	\caption{Shock Hugoniots of aluminum calculated by the HFS model for various ways of choosing $\epsilon_0$. The ionic contribution was taken into account in the approximation of CHS. Red curve---$\epsilon_0 = \epsilon_0(N_\mathrm{max})$, grey curve---$\epsilon_0 = \epsilon_0(N=4)$, black solid curve---$\epsilon_0 = \epsilon_0(N=5)$. Black dashed curve---calculation \cite{Nikiforov:1990}. Experimental data: see figure~\ref{fig2}.} 
\label{fig7}
\end{figure}

The shock Hugoniots of aluminum calculated for different values of $N$ are shown in figure~\ref{fig7}.

\section{Conclusion}
From among the three considered models of electronic part, the TF model provides the worst agreement with the experimental data and only reproduces qualitative behavior of the shock Hugoniot. To obtain appropriate quantitative estimates for the thermodynamic functions of aluminum at shock compression, it is necessary to use models that are more complex.

At pressures $P>2$ TPa, taking into account the experimental data spread and distinctions between different models of ionic contribution, it is possible to assume that the TFC model does not contradict with the experimental data.

The HFS model is more complex. It takes into account shell effects, which are manifested themselves as oscillations associated with the ionization of matter with increasing temperature. In the range of influence of these effects on the position of the shock Hugoniot for solid aluminum samples ($P > 2$ TPa), available experimental data do not provide for observing these oscillations due to insufficient accuracy.

Overall, the HFS model with the CHS model of ionic contribution provides for the best agreement with shock-wave data. The region of particular interest is at pressures $P\approx 0.2\textup{--}1$ TPa (on the shock Hugoniot curve, it corresponds to temperatures $1\textup{--}5$ eV), where the experimental position of the shock Hugoniot of aluminum is determined sufficiently reliably and the results of the HFS model are consistent with the available experimental data. In addition, for this range, the choice of ionic contribution model does not effect on the obtained results. The TFC model is simpler but not consistent with experiment in this range and is not applicable, because the correction to the electronic pressure $\left|\Delta P_{e\/\mathrm{TFC}}/P_{e\/\mathrm{TF}}\right| \approx 0.5\textup{--}1$ is comparable to the value of pressure of the TF model, i.e. the condition of smallness of the corrections is not satisfied.

\ack
The work is financially supported by the Russian Science Foundation grant No.\,14-50-00124.

\section*{References}
\providecommand{\newblock}{}

\end{document}